\documentstyle[12pt,epsf]{article}
\def\bc{\begin{center}}
\def\ec{\end{center}}
\def\be{\begin{equation}}
\def\ee{\end{equation}}
\def\bea{\begin{eqnarray}}
\def\eea{\end{eqnarray}}
\def\ra{\rightarrow}

\def\dg{\dagger}

\def\kg{$K\gamma\!\rightarrow\!~K\pi$}
\def\ds{\displaystyle}
\def\eps{\varepsilon^{\mu\nu\alpha\beta}}

\begin{document}

\renewcommand{\thefootnote}{\fnsymbol{footnote}}

\bc
{\Large \bf The Uses of Chiral Anomaly\\
for Determination of the Number of Colors }\\[4mm]

R. N. Rogalyov\footnote{E-mail: rogalyov@mx.ihep.su}  \\[2mm]
State Research Center of Russia\\
"Institute for High-Energy Physics", Protvino, Russia  \\[9mm]
\ec

\begin{abstract}
\noindent The $N_c$-dependence of the vertices $PPP\gamma$, where $P$ is a pseudoscalar meson
and $N_c$ is the number of colors,
is analyzed with regard for the $N_c$-dependence of the quark charges.
It is shown that the best processes for the determination of $N_c$
are the reactions \kg\ and $\pi^\pm\gamma\ra\pi^\pm\eta$
as well as the decay $\eta\ra\pi^+\pi^-\gamma$.
The measurement of the cross section $\sigma(\pi^-\gamma\ra\pi^-\eta)$ 
at the VES facility at the IHEP agrees with the value $N_c=3$.
\end{abstract}

\setcounter{footnote}{0}
\renewcommand{\thefootnote}{\arabic{footnote} )}

The chiral anomaly \cite{Bardeen} is a fundamental property 
of quantum field theories with chiral fermions
such as the Standard Model (SM). The chiral anomaly
offers a quamtum-mechanical violation of a classical symmetry 
at small distances (electroweak scale) such that its manifestations at large
distances (hadronic scale) are unambiguously determined.
This property distinguishes the chiral anomaly from the
other predictions of the SM; it makes the only effect
of quark--lepton interactions at small distances which
can be described in terms of hadronic fields without 
introducing an additional phenomenological parameter.
For this reason, an experimental study of the chiral
anomaly would be a test of the theoretical foundations
of the elementary particle physics.

Phenomenological implications of the chiral anomaly are 
accounted for by the Wess--Zumino--Witten\footnote{In formula 
(\ref{eq:WZW}) it is assumed that the charge matrix is given by
the expression (\ref{eq:ChMatrix}) for all $N_c$; 
however, it is shown below that this assumption is untrue. 
For this reason, the expression (\ref{eq:WZW}) is valid
only for $N_c=3.$} (WZW) functional
\cite{WessZumino,Witten} 

%\vspace*{-3mm}
\be\label{eq:WZW}
S[U,\ell,r]_{\mbox{\tiny WZW}}^{\tiny (N_c=3)} =  
- \frac{i N_c}{48 \pi^2} \int d^4 x\,\eps
\left\langle U \ell_{\mu} \ell_{\nu} \ell_{\alpha}U^{\dagger} r_{\beta}
+ \frac{1}{4} U \ell_{\mu} U^{\dg} r_{\nu} U \ell_\alpha U^{\dg}
r_{\beta} + \right. 
\ee
$$
+ i U \partial_{\mu} \ell_{\nu} \ell_{\alpha} U^{\dg} r_{\beta} 
+ i \partial_{\mu} r_{\nu} U \ell_{\alpha} U^{\dg} r_{\beta}
- i \Sigma^L_{\mu} \ell_{\nu} U^{\dg} r_{\alpha} U \ell_{\beta}
+ \Sigma^L_{\mu} U^{\dg} \partial_{\nu} r_{\alpha} U \ell_\beta -
$$
$$
- \Sigma^L_{\mu} \Sigma^L_{\nu} U^{\dg} r_{\alpha} U \ell_{\beta}
+ \Sigma^L_{\mu} \ell_{\nu} \partial_{\alpha} \ell_{\beta}
+ \Sigma^L_{\mu} \partial_{\nu} \ell_{\alpha} \ell_{\beta}
- i \Sigma^L_{\mu} \ell_{\nu} \ell_{\alpha} \ell_{\beta} +
$$
$$
\left.  +~\frac{1}{2} \Sigma^L_{\mu} \ell_{\nu} \Sigma^L_{\alpha} \ell_{\beta}
- i \Sigma^L_{\mu} \Sigma^L_{\nu} \Sigma^L_{\alpha} \ell_{\beta}
\right\rangle
 - \left( L \leftrightarrow R \right), 
$$
%\vspace*{-3mm} 
\noindent which should be added to the Lagrangian of the 
chiral perturbation theory. Here  
$N_c$ is the numbetr of colors ($N_c\!\! =\!\! 3$); 
the brackets $\left\langle...\right\rangle$ denote 
trace over flavor indices;
\be\label{eq:SigmaU}
\Sigma^L_\mu = U^{\dg} \partial_\mu U \, ; \qquad
\Sigma^R_\mu = U \partial_\mu U^{\dg} \, ; \qquad
U=\exp \left( {i\Phi \sqrt{2} / F} \right);
\ee
\be\label{eq:ChMatrix}
r_\mu =\ell_\mu = eA_\mu\; Q = A_\mu\;\mbox{diag}\;\left({2\over 3}, -\, {1 \over 3}, -\, {1\over 3} \right);
\ee
$A_\mu$ is the electromagnetic field; $F\!\!=\!\!93$~MeV; symbol
$\left( L \leftrightarrow R \right)$ denotes the substitutions
$U \leftrightarrow U^\dg $, $\ell_\mu \leftrightarrow r_\mu $ and
$\Sigma^L_\mu \leftrightarrow \Sigma^R_\mu $; and 
%\vspace*{3mm}
%\vspace*{3mm}
\begin{displaymath}
 \ \Phi = \left(
\begin{array}{ccc}
\displaystyle {\pi^0 \over \sqrt{2}} + {\eta^8 \over \sqrt{6}}+ {\eta^0 \over \sqrt{3}}
& \pi^+ & K^+ \\[3mm]
 \pi^- & {\displaystyle - {\pi^0 \over \sqrt{2}} + {\eta^8 \over \sqrt{6}} 
+ {\eta^0 \over \sqrt{3}} }& K_0 \\[3mm]
K^- & \bar K_0 & \displaystyle - {2\eta^8 \over \sqrt{6}} + {\eta^0 \over \sqrt{3}}
\end{array}\right).
\end{displaymath}

The functional (\ref{eq:WZW}) determins low-energy behavior
of the amplitudes of the reactions $\pi^0\ra\gamma\gamma$,
$\eta\ra\gamma\gamma$, $\eta\ra\pi^+\pi^-\gamma$, $\pi^+\gamma\ra\pi^+\pi^0$,
$\pi^+\gamma\ra\pi^+\eta$, $K^+\gamma\ra\!K^+\pi^0$ {\it etc}.
Some of these reactions ($\pi^0\ra\gamma\gamma$,
$\eta\ra\gamma\gamma$, $\eta\ra\pi^+\pi^-\gamma$, and $\pi^+\gamma\ra\pi^+\pi^0$)
were used for the determination of the number of colors $N_c$.

However, a recent analysis \cite{Wiese} of the vertices 
$PPP\gamma$ and $P\gamma\gamma$ ($P$ is a pseudoscalar meson)
has revealed that, in a self-consistent theory, the vertices 
$\pi^0\gamma\gamma$ and $\pi^0\pi^+\pi^-\gamma$
are independent of $N_c$, in spite of the fact that,
in any textbook on elementary particle physics (see, for example, \cite{Zuber}),
the width of the decay $\pi^0\ra \gamma\gamma$  is said
to be proportional to $N_c^2$ and thus the width 
$\Gamma(\pi^0\ra \gamma\gamma)$ is considered to be an important
source of experimental information on the value of $N_c$.
The point is that the statement on the dependence of the amplitudes 
${\cal A}_{\pi^+\gamma\ra\pi^+\pi^0}$ and  ${\cal A}_{\pi^0\ra\gamma\gamma}$
on $N_c$ stems from an impicit (and faulty) assumption
that the quark charges $Q_u=2/3, Q_d=-1/3, Q_s=-1/3$ are independent of $N_c$.
If this assumption were true, the triangle anomalies in the quark sector
do not cancel those in the lepton sector and thus the SM 
is not renormalizable. Assuming renormalizability of the SM
for all $N_c$, we obtain the relations between $N_c$ and the quark
charges 
\be\label{eq:QChNcDependent}
Q_u\;=\;{1\over 2}\left({1\over N_c}\,+\,1\right),\ \ \
Q_d\;=\;{1\over 2}\left({1\over N_c}\,-\,1\right).
\ee
Using these relations as the base one can show that the amplitudes 
of the reactions $\pi^0\ra\gamma\gamma,\ \pi^+\gamma\ra\pi^+\pi^0$, and
$\eta\ra\gamma\gamma$ are independent of $N_c$.
The anomalous vertices 
$\gamma\pi^0\pi^+\pi^-$ and $\gamma\eta\pi^+\pi^-$ have been studied
theoretically (in the case $N_c=3$) in 
\cite{Holstein2} and experimentally in the processes of Coulomb production
of $\pi^0$ \cite{Antipov} and $\eta$ \cite{Zaitsev} mesons on nuclei at the IHEP.
It should be mentioned that the motivation of the experiment \cite{Antipov}
(the measurement of the cross section $\sigma(\pi^+\gamma\ra\pi^+\pi^0)$) was 
to determine the number of colors; however, according to the above,
the number of colors could not be determined in this experiment.
Therewith, the data obtained in the experiment \cite{Zaitsev} can well
be used for a determination of $N_c$.
As for now, the only vertex involving light mesons used for a
determination of the number of colors $N_c$ is $\eta\pi^+\pi^-\gamma$.
It has been studied in the decay $\eta\ra\pi^+\pi^-\gamma$ \cite{EtaPiPiGammaExp}
and in the scattering process $\pi^+\gamma\ra\pi^+\eta$ \cite{Zaitsev}.
The expression for this vertex for an arbitrary value of $N_c$
is presented below. With regard for this expression, the mentioned
experiments give evidence for the value $N_c=3$.

The present study as well as \cite{Wiese}
does not cast doubt on the total of the experimental data 
used for the determination of the parameter $N_c$;
we pursue rather unpretentious goals.
B\" ar and Wiese \cite{Wiese} suggest that the  
decay $\eta\ra\pi^+\pi^-\gamma$ "should replace 
the textbook example $\pi^0 \rightarrow \gamma \gamma$ for lending experimental
support to the fact that there are three colors in our world."
In the present study, we consider the $N_c$-dependence of
the cross sections of the reactions \kg\ and $K\gamma\ra~K\eta$,
which can also be used for the determination of $N_c$.
A detailed study of the vertices $KK\pi\gamma$ and $KK\eta\gamma$
is needed because, in spite of a sophisticated and comprehensive
analysis of the $P\gamma\gamma$ and $PPP\gamma$ vertices
performed in  \cite{Wiese}, the ultimate expressions for 
these vertices (formula (5.11) of the cited paper)
are in error. For istance, the mentioned formula in the case $N_c=3$
disagree with that obtained from the WZW Lagrangian  (\ref{eq:WZW}).

The effective Lagrangian for the vertices $PPP\gamma$
can be calculated by either of two ways:
\begin{enumerate}
\item By a straightforward computation of the group-theoretical 
coefficients of the quark digarams contributing to 
the respective Green function with antisymmetrization over 
axial currents and regard for the relations (\ref{eq:QChNcDependent}). 
Strictly speaking, this procedure yields only the ratios
between the coefficients of the same type; the known vertices
 $\pi^0\gamma\gamma$ and  $\pi^+\pi^-\pi^0\gamma$ can be used as the
 reference values.
\item By a substitution of the explicit expression for the matrix $U$ (formula (\ref{eq:SigmaU}))
in the expression for the  WZW Lagrangian (\ref{eq:WZW}), generalized
to the case $N_c\neq 3$. Such generalization was proposed in \cite{Wiese}.
It has the form
\be
S=S^{\tiny (N_c=3)}+\left(1-{N_c\over 3}\right)\;S_{\mbox{\tiny GW}},
\ee
where $\ds S_{\mbox{\tiny GW}}$ is the Goldstone--Wilczek current \cite{GW}: 
\bea\label{eq:GWcurrent}
S_{\mbox{\tiny GW}}[U,A_\mu]&=&\frac{e}{48 \pi^2} \int d^4x \ \eps A_\mu \mbox{Tr}
[(U^\dagger \partial_\nu U)(U^\dagger \partial_\alpha U)(U^\dagger \partial_\beta U)] \nonumber \\
&-&\frac{i e^2}{32 \pi^2} \int d^4x \ \eps A_\mu F_{\nu\alpha} 
\mbox{Tr}[Q(\partial_\beta U U^\dagger + U^\dagger \partial_\sigma U)];
\eea
\end{enumerate}
\vspace*{-1mm}
when considering the $PPP\gamma$ vertices, the second row in 
formula (\ref{eq:GWcurrent}) can be omitted.

The results of the calculations by these two ways coincide;
the vertices sought for have the form
\bea\label{eq:NcDependencePPPgamma}
{\cal L}^{PPP\gamma}_{\mbox{\tiny WZW}} = 
{ie \over 4\pi^2F^3}\varepsilon^{\mu\nu\alpha\beta} A_\beta 
 \hspace*{-2mm} &\ & \hspace*{-8mm} \left(\hspace*{23mm}  \partial_\mu \pi^0 \partial_\nu \pi^+  \partial_\alpha \pi^-  \right. \; + \\ \nonumber
\hspace*{-2mm} &+& %\hspace*{-2mm} 
\ {N_c \over 3\sqrt{3}}\hspace*{8mm} 	\partial_\mu \eta^8 \partial_\nu \pi^+ \partial_\alpha \pi^-   \; + \\ \nonumber
\hspace*{-2mm} &+& %\hspace*{-2mm} 
\ {N_c +3 \over 6}\hspace*{4mm}	\partial_\mu \pi^0  \partial_\nu   K^+ \partial_\alpha K^-   \; + \\ \nonumber
\hspace*{-2mm} &+& %\hspace*{-2mm} 
\ {N_c -1 \over 2}\hspace*{4mm}\partial_\mu \pi^0  \partial_\nu K^0   \partial_\alpha \bar K^0   \; + \\ \nonumber
\hspace*{-2mm} &+& %\hspace*{-2mm} 
\ {9- N_c  \over 6\sqrt{3}}\hspace*{3mm} \partial_\mu \eta^8 \partial_\nu K^+   \partial_\alpha K^-   \; - \\ \nonumber
\hspace*{-2mm} &-& \hspace*{-3mm} 
\ {\sqrt{3}(N_c -1) \over 2}\hspace*{1mm} 	\partial_\mu \eta^8 \partial_\nu K^0   \partial_\alpha \bar K^0   \; - \\ \nonumber
\hspace*{-2mm} &-& %\hspace*{-2mm} 
\ {N_c -3 \over 3\sqrt{2}}\hspace*{2mm} 	\partial_\mu \pi^-  \partial_\nu K^+   \partial_\alpha \bar K^0   \; + \\ \nonumber
\hspace*{-2mm} &+& %\hspace*{-2mm} 
\ {N_c -3 \over 3\sqrt{2}}\hspace*{2mm}	\partial_\mu \pi^+  \partial_\nu K^-   \partial_\alpha K^0   \; + \\ \nonumber
\hspace*{-2mm} &+& %\hspace*{-2mm} 
\ {\sqrt{6} \over 9}\hspace*{8mm}	\partial_\mu \eta^0 \partial_\nu K^+   \partial_\alpha K^-   \; + \\ \nonumber
 \hspace*{-2mm} &+&  \left. \ {N_c \sqrt{6} \over 9}\hspace*{4mm} \partial_\mu \eta^0 \partial_\nu \pi^+  \partial_\alpha \pi^-  \right),  \nonumber
\eea
where $\eta^0$ and $\eta^8$ are the singlet and octet states:
\be\label{eq:mixing} 
\eta =\eta^8 \cos\theta_P - \eta^0 \sin\theta_P, \hspace*{12mm} \eta' = 
\eta^8 \sin\theta_P + \eta^0 \cos\theta_P,  \hspace*{12mm} \theta_P\simeq 20^o.
\ee
It should be noticed that the vertices $K^+K^0\pi^-\gamma$ and $K^-\bar K^0\pi^+\gamma$ 
do not appear in the anomalous action\footnote{
The vertices $K^+K^0\pi^-\gamma$ and $K^-\bar K^0\pi^+\gamma$ 
 are not presented in \cite{Wiese}.}
only in the case $N_c=3$. For this reason, the near-threshold behavior
of the reactions $K^+\gamma\ra\!K^0\pi^+$ and $K^0\gamma\ra\!K^+\pi^-$ 
offer a good indicator of a deviation of the parameter $N_c$ from the value $N_c=3$. 
To put it differently, the chiral anomaly gives a contribution 
to the amplitudes of the reactions $K^+\gamma \ra K^+\pi^0$ and $K^0\gamma \ra K^0\pi^0$
and gives no contribution to the amplitudes of the reactions
$K^+\gamma \ra K^0\pi^+$ and $K^0\gamma \ra K^+\pi^-$
{\bf only at $N_c=3$}. As a consequence of this,
the cross sections of the reactions 
$K^+\gamma \ra K^+\pi^0$ and $K^0\gamma \ra K^0\pi^0$
over the near-threshold region far exceed the cross sections of the reactions 
$K^+\gamma \ra K^0\pi^+$ and $K^0\gamma \ra K^+\pi^-$.
These cross sections were calculated {\bf (in the case $N_c=3$)}
in \cite{RogYAF2001}, where a possibility of an experimental study
of these cross sections is discussed.
A measurement of the cross sections of the reactions
$K^+\gamma \ra K^+\pi^0$, $K^0\gamma \ra K^0\pi^0$,
$K^+\gamma \ra K^0\pi^+$, and $K^0\gamma \ra K^+\pi^-$
is of particular interest due to their dependence on $N_c$, 
ptresented in formula (\ref{eq:NcDependencePPPgamma}).

However, the formilas (\ref{eq:NcDependencePPPgamma}) give an adequate
description of the amplitudes only at sufficiently small momenta.
To describe the reactions \kg, $\pi^\pm\gamma\ra\pi^\pm\eta$
and the decay $\eta\ra\pi^+\pi^-\gamma$ at the physical values of momenta,
one should take into account the contribution of the $1^{-\;-}$ resonances,
which can be calculated in the vector-meson dominance model.
In what follows, we use the version of the vector-meson dominance model
proposed in \cite{Bando}, because {\it (i)} in the chiral limit,
it goes over into the chiral perturbation theory; {\it (ii)}
the formalism proposed in \cite{Bando} is well suited for 
taking the chiral anomaly into consideration.
The Lagrangian of this model and its application to the above-mentioned
processes can be found in \cite{RogYAF2001, Bando, Fujiwara}; here we
only point to its dependence on $N_c$.
All $N_c$-dependence in the normal part of this Lagrangian is absorbed in 
the effective coupling $g$. The anomalous terms should be mutiplied by
$\ds {N_c\over 3}$ in order to obtain the WZW Lagrangian in the vicinity of 
the chiral limit; the quark charges are considered to be the functions of $N_c$, 
according to (\ref{eq:QChNcDependent}).
The analysis of the reactions $\pi^\pm\gamma\ra\pi^\pm\eta$ and the decay
$\eta\ra\pi^+\pi^-\gamma$ in the vector-meson dominance model
was performed in \cite{Holstein2}.
From the expressions presented in \cite{Holstein2} it follows that
both diagrams for these processes are proportional to $N_c(Q_u-Q_d)=N_c$
and the dependence of the amplitudes on $N_c$ is readily determined.

The reactions \kg\ present a challenge: a straightforward
computation of the amplitudes is needed. The respective 
Feynman diagrams are presented in the figure.

\begin{figure}%\vspace*{15mm}
\hbox{
       \epsfxsize=477pt \epsfbox{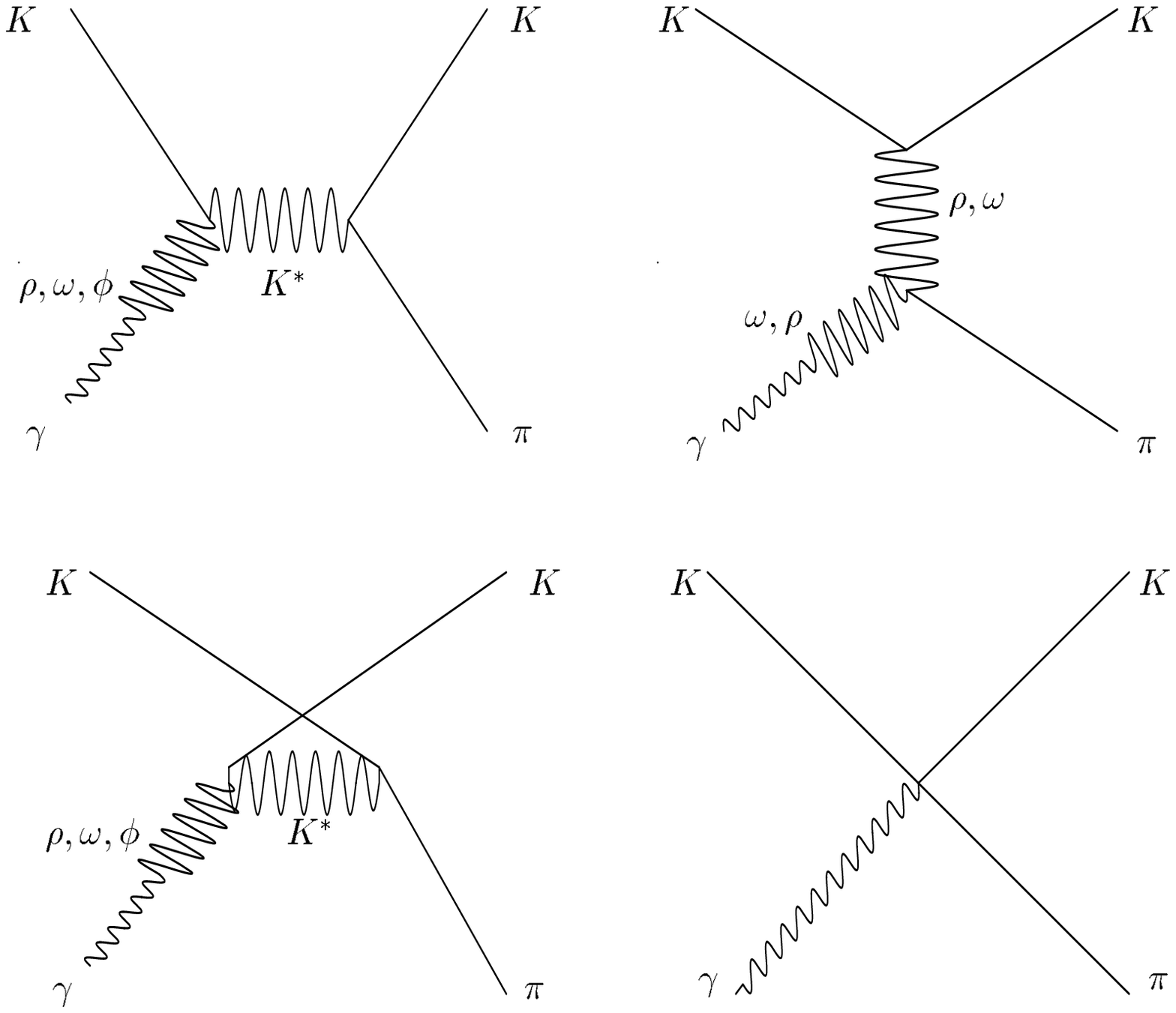} \hspace*{10pt}
}
\caption{ Diagrams  contributing to the reactions \kg.
}\label{fig:KgKPVMD}
\end{figure}
$\ $

The calculations were performed with the FORM package \cite{FORM},
the result has the form
\bea\label{eq:ScAmpl}
A_{K\gamma \ra K\pi} =  {- ie \over 16\pi^2F^3} 
\eps q^\mu p_b^\nu p_2^\alpha \epsilon^\beta 
\left( C_0  + {C_s M_{K^\ast}^2 \over s - 
M_{K^\ast}^2 + i\Gamma_{K^\ast}\sqrt{s}}\; + \right. \\[1mm] \nonumber
+ \left. {C_t M_{\rho}^2 \over t - M_{\rho}^2 } +
{C_u M_{K^\ast}^2 \over u - M_{K^\ast}^2 } \right), \nonumber
\eea
where $\epsilon$ is the photon polarization vector; 
the momenta $q$, $p_b$, and $p_2$ are defined in the table; 
$s=(q+p_b)^2,\ t=(p_b-p_2)^2,\ u=(q-p_2)^2$;
$M_{K^\ast}$ and $M_\rho$ are the masses of the $K^\ast$ and $\rho$ mesons; 
$\Gamma_{K^\ast}$ is the width of the $K^\ast$ meson.
The coefficients $C_0$, $C_s$, $C_t$, and $C_u$ for specific
processes are presented in the table as the functions of $N_c$.

\vspace*{4mm}
{\bf Table.} Coefficients $C_s, C_t, C_u$, and $C_0$ from formula (\ref{eq:ScAmpl}).\\[1mm]

\begin{tabular}{|c|c|c|c|c|}\hline
&&&&\\
Reaction & $C_0$ & $C_s$ & $C_t$ & $C_u$ \\ 
&&&&\\
\hline
&&&&\\
$K^+(p_b)\gamma(q) \ra K^+(p_2)\pi^0(p_1)$ & $\ds {N_c+3 \over 3}$ & \hspace*{3mm} 1 \hspace*{3mm} & $N_c+1$ & \hspace*{3mm} 1 \hspace*{3mm} \\ 
&&&&\\
\hline
&&&&\\
$K^+(p_b)\gamma(q) \ra K^0(p_2)\pi^+(p_1)$ & $\ds {\sqrt{2}(N_c-3)\over 3}$ & $\sqrt{2}$ & $\sqrt{2} (N_c-2)$ & $-2 \sqrt{2}$ \\ 
&&&&\\
\hline
&&&&\\
$\pi^+(p_b)\gamma(q) \ra \pi^+(p_2)\eta(p_1)$ & $\ds {2N_c\sqrt{3}\over 9}\;P_\theta$ & 0 & $\ds {2N_c\sqrt{3}\over 3}\;P_\theta$ & 0 \\ 
&&&&\\
\hline
\end{tabular}
\vspace*{4mm}
\bc
\begin{minipage}{0.8\hsize}
{\small {\bf Note.} Taking into account the $\eta-\eta'$ mixing (\ref{eq:mixing}) 
gives rise to the factor \cite{Holstein2} 
$\ds P_\theta=\frac{F}{F_8}\,\cos\theta_P-\frac{F}{F_0}\,\sin\theta_P$, 
where the difference between the decay constants $F_{\pi^\pm}=F$, $F_0$, and $F_8$
must also be taken into consideration.
The values $F_0$ ($F_8$) parametrize the matrix element of the axial current
between vacuum and purely singlet (octet) state.
Their magnites were computed in the one-loop approximation of the
chiral perturbation theory and are equal to \cite{Gasser}:
$F_0\approx 1.04 F$, $F_8\approx 1.30 F$; the mixing angle is $\theta_P=20^o$.
All this makes theoretical predictions for the decays $\eta\ra\pi^+\pi^-\gamma$
and $\eta'\ra\pi^+\pi^-\gamma$ and the reactions $\pi^\pm\gamma\ra\pi^\pm\eta(\eta')$ 
less accurate than those for the reactions \kg.}
\end{minipage}
\ec

Thus a measurement of the cross sections of the reactions
$K^+\gamma \ra K^+\pi^0$ and $K^+\gamma \ra K^0\pi^+$
would compensate for a deficiency in experimental facts
giving evidence for $N_c=3$.
This deficiency is due to an exclusion of the decays
$\pi^0 \ra \gamma\gamma$ and $\eta \ra \gamma\gamma$ and the reaction 
$\pi^+\gamma\ra\pi^+\pi^0$ from the total of the experimental
data used for a determination of $N_c$.
Moreover, a measurement of the cross sections of the reactions \kg\ 
would make it possible to check the phenomenological
implications of the chiral anomaly in the world with three
(rather than two) light quarks.
It is important because the WZW Lagrangian was derived 
under the assumption that there are just three light quarks.

I am grateful to M.I.Polykarpov for the interest to this study.

\end{document}